\documentclass[a4paper,aps,prd,twocolumn,groupedaddress,amssymb,eqsecnum,showpacs,nofootinbib]{revtex4}
\usepackage{graphicx}
\usepackage{mathptmx}
\usepackage{bm}
\usepackage{times}

\begin{document}

%\preprint{}

\title{Trans-Planckian relics in the scalar to tensor ratio}

\author{Hael Collins}
\email{hael@nbi.dk}
\affiliation{The Niels Bohr International Academy, The Niels Bohr Institute, 2100 Copenhagen \O, Denmark}

\date{\today}

\begin{abstract}
The physical properties of our universe at energy scales above the expansion rate during inflation can affect predictions for the ratio between the amplitudes of the primordial scalar and tensor fluctuations.  In particular, we study here the effects of a breakdown of a locally Lorentz invariant description of nature at tiny space-time intervals.  In some instances, these effects shift the amplitudes by a constant amount, altering the standard relation between this ratio and the slow-roll parameters.  More generally, ``trans-Planckian'' effects introduce a modulation in the primordial power spectra which grows at shorter scales, making the value of the ratio sensitive to the scale at which it is defined.  We also present a model where symmetries are broken at horizon scales during inflation.  In this case, the power at large scales today could then be suppressed, relative to that at smaller scales.
\end{abstract}


\maketitle

\section{Introduction}
\label{intro}

In our current picture of the universe, the bewildering variety of structures that we see throughout the universe all began as tiny inhomogeneities in an otherwise featureless space-time.  These tiny ripples in the background are perhaps the most ancient relics left from the very early universe whose direct influence we can still observe.  They appear to be present at every scale we can see, including the very largest---there are fluctuations whose spatial extent is comparable to the size of the observable universe today.

In inflation, these ripples are made through two simple ingredients:  the quantum fluctuations of the metric and an expanding space-time.  The metric along with the fields driving the inflationary expansion are each divided into a classical part plus a tiny quantum piece.  The classical component describes a spatially uniform expanding universe.  The second piece, being a quantum field, is always fluctuating; it thereby introduces a small spatial variation into an otherwise featureless universe.  The spatial extent of any one of these fluctuations does not stay fixed, since its size is stretched ever larger and larger as the universe expands.  And since the rate of this expansion is {\it accelerating\/} during inflation, a fluctuation is eventually stretched to a size beyond the range of any later causal process, at least while the inflationary era lasts.  At that point it remains essentially frozen into the background, still passively growing as the universe continues to expand.  When inflation comes to an end, the range of causal processes begins to overtake the rate at which the fluctuations are growing.  Once overtaken, they can then influence the evolution of the other ingredients of the universe.  They effectively become the initial inhomogeneities needed to explain the origin of all the rich structure of our cosmos.

In essence then, inflation relies on the quantum fluctuations of gravity.  By itself, this situation is not immediately dangerous.  As long as we never consider variations of the metric on scales smaller than a Planck length ($\ell_{\rm pl} \sim\sqrt{8\pi G}$, where $G$ is the Newton constant), quantum gravity has a perfectly sensible perturbative description.  However, in a time-evolving background, whether a fluctuation is larger or smaller than some fixed scale depends on {\it when\/} we are looking at it.  A fluctuation that is larger than a Planck length at one moment might have been smaller than it just a little while before.

It turns out that the allowed range of fluctuations that avoid this Planckian regime is quite constrained---perhaps even nonexistent.  To put it a little more precisely, if we define $\lambda_{\rm max}$ to be the length of the largest causally consistent fluctuation possible at the start of inflation and $H$ to be the expansion rate of the universe during inflation, then the range of safe fluctuations are those whose wavelengths at the beginning of inflation lie within the range
\begin{equation}
(H\ell_{\rm pl})\, \lambda_{\rm max} < \lambda < \lambda_{\rm max} , 
\label{lambdarange}
\end{equation}
that is, only $\log(H\ell_{\rm pl})^{-1}$ orders of magnitude.  This range is actually the best possible case.  It is realized only when the scale we have called $\lambda_{\rm max}$ grows until it is precisely the size of the observable universe today.  Since this scale is just the largest causally allowed scale when inflation begins, we could well have had a bit more inflation than this absolutely minimal amount, in which case the range of allowed wavelengths would then be narrower.  In fact, if inflation lasts just $\ln(H\ell_{\rm pl})^{-1}$ $e$-folds longer  than the minimal case, then this range disappears altogether and every scale whose influence we see today would have been once smaller than a Planck length.

This reliance of inflation on what happens at scales where we have no adequate understanding of nature constitutes the {\it trans-Planckian problem of inflation\/} \cite{brandenberger}.  Thus far, most of the work on this problem has not so much attempted to resolve it as it has tried to describe how some specific new property of nature near the Planck scale would appear in experiments.  The best places to look for these effects are in those features of the universe that are the most directly influenced by the original pattern of the primordial fluctuations produced by inflation.  In practice, these places are the cosmic microwave background and the distribution of matter over vast regions of the universe.

Since the perturbations generated by inflation are due to fluctuations in the metric, they can be classified and treated separately according to their {\it spatial\/} symmetry properties.  The most important---since their influence on the local matter density is the strongest---are the scalar perturbations.  Yet inflation predicts the existence of other sorts as well.  The quantum fluctuations of the metric during the inflationary epoch also produce a bath of spin-two tensor fluctuations---gravity waves---throughout the universe.  

Evidence for the scalar perturbations has been seen in many experiments, but so far the tensor perturbations have been more elusive.  However, the {\it ratio\/} of the amplitudes of these fluctuations can have observable consequences and can thereby be constrained, even if the direct gravitational wave signal has not been measured.  How this ratio enters into various inflationary predictions depends critically on what assumptions we are making at trans-Planckian scales, so it is important to understand these assumptions---including those of the standard picture too---if we are making inferences about the gravity waves without having seen them directly.

This article explores some of the ways potential ``trans-Planckian'' effects would alter our expectations for the scalar to tensor ratio.  In particular, we shall examine what happens when space-time is not perfectly flat at infinitesimal distances, but breaks some of the flat-space symmetries.  Such broken symmetries can substantially enhance or diminish this ratio, even while having a relatively minor effect on the scalar or tensor perturbations considered separately.  We shall also discover that the amplitude of the scalar or tensor fluctuations can depend much more sensitively on the scale at which it is defined than is usually thought.  Trans-Planckian corrections often introduce oscillations into the amplitudes, unlike the very weak power-law dependence of the standard prediction.

Because the trans-Planckian problem arises so naturally out of the basic ingredients of inflation, we describe the standard picture for how the primordial perturbations are generated in inflation in Sec.~\ref{prim}.  The conditions generally assumed for an inflationary universe mean that it closely resembles de Sitter space, a universe with a constant vacuum energy density.  Therefore, in Sec.~\ref{deSitter} we give an example of how one of the symmetry-breaking effects in this simpler setting would alter our expectations for the size of the scalar to tensor ratio.  Section \ref{slow} describes the general signatures of the trans-Planckian symmetry-breaking effects in a genuinely slowly rolling universe, which are then discussed in a series of illustrative cases in the following section.  The final section concludes with a few more general observations.

\section{The primordial perturbations}
\label{prim}

\subsection{The quadratic action}

Let us begin by recalling how the primordial perturbations are generated in inflation.  We consider a fairly simple picture, one with only gravity and a single scalar inflaton, $\phi$, where the dynamics are described by the sum of the gravitational action
\begin{equation}
S_g = {1\over 16\pi G} \int d^4x\, \sqrt{-g}\, R
\label{gaction}
\end{equation}
and the action for the inflaton, 
\begin{equation}
S_\phi = \int d^4x\, \sqrt{-g}\, \bigl[ 
{\textstyle{1\over 2}} \nabla_\lambda \phi \nabla^\lambda \phi
- V(\phi) \bigr] . 
\label{phiaction}
\end{equation}
Any cosmological constant can be implicitly included in the inflaton's potential, $V(\phi)$, so we have not written it explicitly in the gravitational action.

Since the universe appears to be globally flat and---at least at its early stages---extremely uniform, we shall use a conformally flat metric, 
\begin{equation}
ds^2 = a^2(\eta)\, \bigl[ d\eta^2 - d\vec x\cdot d\vec x \bigr] , 
\label{metric}
\end{equation}
along with tiny quantum fluctuations about this background.  The rate at which the scale factor $a(\eta)$ changes defines a natural dynamical scale associated with this geometry, 
\begin{equation}
H = {a'\over a^2} \equiv {1\over a^2} {\partial a\over\partial\eta} , 
\label{Hubbledef}
\end{equation}
called the Hubble scale.

Both the metric and the inflaton are expected to undergo quantum fluctuations about classical, spatially constant, values
\begin{eqnarray} 
g_{\mu\nu}(\eta,\vec x) 
&\!\!\!=\!\!\!& a^2(\eta)\, \eta_{\mu\nu} + \delta g_{\mu\nu}(\eta,\vec x) 
\nonumber \\
\phi(\eta,\vec x) 
&\!\!\!=\!\!\!& \phi_0(\eta) + \delta\phi(\eta,\vec x) , 
\label{flucts}
\end{eqnarray}
where $\eta_{\mu\nu}$ is the Minkowski space metric.  Although the time-dependence in the leading part of the metric, $a^2(\eta)\eta_{\mu\nu}$, generally means that the space-time has fewer symmetries than Minkowski space (where $a=1$) or de Sitter space (where $a=-(H\eta)^{-1}$), the {\it spatial\/} symmetries remain unbroken.  We can thus characterize the quantum fluctuations by how they transform with respect to this spatial symmetry; it also ensures that different classes of fluctuations do not influence each other at leading order.  

The most important fluctuations phenomenologically are those which transform as scalars and as spin-two tensors under spatial rotations.  The quantum fluctuations of the inflaton also transform as a scalar and therefore they can mix with the scalar component of the metric fluctuations.  A completely general set of scalar fluctuations is given by 
\begin{eqnarray} 
&&\!\!\!\!\!\!\!\!\!\!\!\!\!\!\!\!\!
\delta g_{\mu\nu}\, dx^\mu dx^\nu  
\nonumber \\
&\!\!\!=\!\!\!& 2 a^2(\eta)\, \Bigl[ 
\bigl( \Phi - a^{-1}[(B-E')a]' \bigr)\, d\eta^2 
- \partial_i B\, d\eta dx^i
\nonumber \\
&&\qquad\quad
+ \bigl( (\Psi+aH(B-E'))\delta_{ij} - \partial_i\partial_j E \bigr)\, dx^i dx^j 
\Bigr] \quad 
\label{scalarA}
\end{eqnarray}
for the metric and 
\begin{equation}
\delta\phi = \delta\varphi - \phi'_0 (B-E') 
\label{scalarB}
\end{equation}
for the inflaton.  The reason for this complicated form, with its recurring appearances of $B-E'$, is that this form leaves the fields $\Phi$, $\Psi$ and $\delta\varphi$ unchanged when we slightly transform our coordinate system.  The fields $\Phi$ and $\Psi$ in particular were introduced by Bardeen in \cite{bardeen}.

To determine the basic dynamics of these scalar perturbations, we must expand the combined action, $S_g+S_\phi$, and then extract its quadratic terms.  This calculation is quite lengthy and is already described in full in \cite{cospert}, so we shall only mention its final results here.  In the course of expanding the action, a constraint arises among the three coordinate-invariant fields, 
\begin{equation}
\Psi' + aH\Phi = 4\pi G\, \phi'_0 \delta\varphi . 
\label{scalarconst}
\end{equation}
If we apply this constraint and introduce $\varphi$, a linear combination of the inflaton and the scalar part of the metric, defined by 
\begin{equation}
\varphi = \delta\varphi + {\phi'_0\over aH} \Psi , 
\label{varphidef}
\end{equation}
we discover in the end a simple action for the scalar perturbations, given in terms of only a single dynamical field, $\varphi$,
\begin{equation}
S^{(2)}_\varphi = \int d^4x\, a^2\, \Bigl[ 
{\textstyle{1\over 2}} \eta^{\mu\nu} \partial_\mu\varphi \partial_\nu\varphi 
- {\textstyle{1\over 2}} a^2m^2\, \varphi^2 
\Bigr] . 
\label{scalaraction}
\end{equation}
The effective mass of this field, $m^2$, inherits its time-dependence both from the background and from the inflaton's potential $V(\phi)$,
\begin{equation}
m^2 = {\delta^2 V\over \delta\phi_0^2} + {4\over a}H' 
+ {2\over a^2} {H^{\prime\prime}\over H} 
- {2\over a^2} {H^{\prime 2}\over H^2} . 
\label{massdef}
\end{equation}
The other degrees of freedom---the linear combinations orthogonal to $\varphi$ as well as $B$ and $E$---only appear as total derivatives.  Though these much lengthier terms are also quadratic in the fields, they have no dynamical effect \cite{cospert}.

The treatment of the tensor fluctuations is much more straightforward since the metric is typically the only available source and moreover they are already invariant under small coordinate transformations,
\begin{equation}
\delta g_{\mu\nu} = a^2(\eta)\, h_{ij}(\eta,\vec x)\, dx^i dx^j . 
\label{tensordef}
\end{equation}
Here $h_{ij}$ is traceless and transverse, 
\begin{equation}
h_i^i = 0 \qquad \partial_i h_j^i = 0 , 
\label{transtrace}
\end{equation}
so that it does not contain any scalar or vector components.  Expanding the action again and this time isolating its quadratic terms in $h_{ij}$ yields 
\begin{equation}
S^{(2)}_h = {1\over 64\pi G} \int d^4x\, a^2\bigl[ 
\eta^{\mu\nu} \partial_\mu h_{ij} \partial_\nu h^{ij} 
\bigr] . 
\label{tensoraction}
\end{equation}
To give the field a more canonical normalization, we can introduce 
\begin{equation}
\tau_{ij} = {1\over\sqrt{16\pi G}} h_{ij} , 
\label{taudef}
\end{equation}
to obtain
\begin{equation}
S^{(2)}_\tau = \int d^4x\, a^2 \Bigl[ 
{\textstyle{1\over 4}} \eta^{\mu\nu} \partial_\mu\tau_{ij} \partial_\nu\tau^{ij} 
\Bigr] . 
\label{tauaction}
\end{equation}

These two quadratic actions $S^{(2)}_\varphi$ and $S^{(2)}_\tau$ determine the behavior of the quantum fluctuations of the inflaton along with the background space-time.  Once we have solved for this behavior, together with some assumptions about the initial states of these fields, we can describe the patterns of these perturbations through their correlation functions.

\subsection{The power spectrum}

The most easily observed pattern in these primordial perturbations is the two-point function, which tells how the perturbations in two different places are correlated.  For the case of the scalar $\varphi$ and the tensor $h_{ij}$ fluctuations we have
\begin{eqnarray}
&\langle 0(\eta) | \varphi(\eta,\vec x)\varphi(\eta, \vec y) | 0(\eta)\rangle &
\nonumber \\
&\langle 0(\eta) | h_{ij}(\eta,\vec x) h^{ij}(\eta, \vec y) | 0(\eta)\rangle . & 
\label{twopoints}
\end{eqnarray}
In writing the matrix elements thus, we are tacitly working in the interaction picture.  There, the evolution of an operator---for example, the product of the field at two different points---is determined by the free part of the Hamiltonian, while the evolution of the states is governed by its interacting part, $H_I(\eta)$.  To leading order, where we neglect the influence of terms other than those already included in $S^{(2)}_\varphi$ and $S^{(2)}_\tau$, we can also neglect the time-evolution of the states once we have established it at some initial time $\eta_0$, 
\begin{equation}
|0(\eta)\rangle \to |0\rangle \equiv |0(\eta_0)\rangle . 
\label{vacinit}
\end{equation}
Although we have written the vacuum as $|0(\eta)\rangle$ for both the scalar and tensor modes, to be more precise we need to define the vacuum state separately for each of these fields, along with any others that might be present, 
\begin{equation}
|0(\eta)\rangle = |0(\eta)\rangle_\varphi \otimes |0(\eta)\rangle_h \otimes \cdots ; 
\label{fullvac}
\end{equation}
but to keep the notation simple we shall continue to write the state as just $|0(\eta)\rangle$.  Nevertheless, the vacuum can in principle be defined differently for the scalar and the tensor modes, and it might even be that one or the other of the fields is not even in its vacuum state at all, as could happen when one happens to be coupled to some other excited field \cite{cliff}.

A two-point function is more easily analyzed in terms of its Fourier transform, called the {\it power spectrum\/}, which we accordingly define for both the scalar, $P_k^\varphi(\eta)$, and tensor, $P_k^h(\eta)$, modes by
\begin{equation}
\langle 0(\eta)| \varphi(\eta,\vec x)\varphi(\eta,\vec y) |0(\eta)\rangle 
= \int {d^3\vec k\over (2\pi)^3}\, e^{i\vec k\cdot(\vec x-\vec y)} 
{2\pi^2\over k^3} P_k^\varphi(\eta)
\nonumber \\
\label{phipower}
\end{equation}
and
\begin{equation}
\langle 0(\eta)| h_{ij}(\eta,\vec x) h^{ij}(\eta,\vec y) |0(\eta)\rangle 
= \int {d^3\vec k\over (2\pi)^3}\, e^{i\vec k\cdot(\vec x-\vec y)} 
{2\pi^2\over k^3} P_k^h(\eta) . 
\label{hpower}
\end{equation}
Note that $k\equiv |\vec k|$.  We similarly expand the scalar and tensor fields in creation and annihilation operators, together with their associated time-dependent eigenmodes, as
\begin{equation}
\varphi(\eta,\vec x) 
= \int {d^3\vec k\over (2\pi)^3}\, \bigl[ 
\varphi_k(\eta) e^{i\vec k\cdot \vec x} a_{\vec k} 
+ \varphi_k^*(\eta) e^{-i\vec k\cdot \vec x} a_{\vec k}^\dagger 
\bigr] 
\label{phiop}
\end{equation}
and 
\begin{eqnarray}
&&\!\!\!\!\!\!\!\!
\tau_{ij}(\eta,\vec x) = 
\nonumber \\
&&\!\!\!\!\!
\int {d^3\vec k\over (2\pi)^3}\, \sum_{r=1}^2 \bigl[ 
\tau_k^{(r)}(\eta) e^{i\vec k\cdot \vec x} e_{\vec k,ij}^{(r)} a_{\vec k}^{(r)} 
+ \tau_k^{(r)*}(\eta) e^{-i\vec k\cdot \vec x} e_{\vec k,ij}^{(r)*} a_{\vec k}^{(r)\dagger} 
\bigr] . 
\nonumber \\
&&
\label{tauop}
\end{eqnarray}
The $e_{\vec k, ij}^{(r)}$ represent the two possible polarizations for the tensor modes.  We have written the annihilation operators for the fields with the same notation, $a_{\vec k}$ or $a_{\vec k}^{(r)}$, since the tensor modes are always accompanied by index for the polarization, so their is little danger of confusing them.  The only non-vanishing commutators are 
\begin{eqnarray}
\bigl[ a_{\vec k}, a_{\vec k'}^\dagger \bigr] 
&\!\!\!=\!\!\!& (2\pi)^3\, \delta^3(\vec k - \vec k') 
\nonumber \\
\bigl[ a_{\vec k}^{(r)}, a_{\vec k'}^{(s)\dagger} \bigr] 
&\!\!\!=\!\!\!& (2\pi)^3\, \delta^{rs} \delta^3(\vec k - \vec k') . 
\label{commute}
\end{eqnarray}

Varying the quadratic terms of the free part of the theory, that is $S^{(2)}_\varphi$ and $S^{(2)}_\tau$, with respect to the appropriate fields, we learn that the time evolution of the modes satisfy the following Klein-Gordon equations, 
\begin{equation}
\varphi_k^{\prime\prime} + 2aH\, \varphi_k^\prime 
+ (k^2 + a^2 m^2) \varphi_k = 0
\label{phikeq}
\end{equation}
and
\begin{equation}
\tau_k^{(r)\prime\prime} + 2aH\, \tau_k^{(r)\prime} 
+ k^2 \tau_k^{(r)} = 0 . 
\label{taukeq}
\end{equation}

\subsection{Slowly rolling inflation}

For a completely general isotropically expanding space-time, it is rarely possible to solve these equations with a simple, analytic expression for the eigenmodes, $\varphi_k$ and $\tau_k$.  However, the inflationary background is something of a special case; we can use some of the constraints already placed on the expansion to obtain a natural method for approximating the solutions for these modes.

A stage of accelerated expansion requires that the potential energy should be significantly larger than the kinetic energy, and for a sufficiently long time that---at the very least---the entire observable part of the universe once lay within a single Hubble horizon, $1/H$.  To put these requirements a little more precisely, if we define the parameters,
\begin{equation}
\epsilon \equiv - {H'\over aH^2}, 
\qquad
\delta + 1 \equiv {1\over aH} {\phi_0^{\prime\prime}\over\phi_0^\prime} ,
\qquad 
\xi \equiv {\epsilon'+\delta'\over aH} . 
\label{slowroll}
\end{equation}
then they translate into the conditions, 
\begin{equation}
\epsilon \ll 1 , 
\qquad
\delta \ll 1 , 
\label{slowrollsmall}
\end{equation}
and $\xi\sim {\cal O}(\epsilon^2, \epsilon\delta, \delta^2)$.  Because the kinetic energy during this era is small, it is often called the slowly rolling regime of inflation.  In terms of these parameters, the effective mass of the scalar modes becomes
\begin{equation}
m^2 = - H^2 \bigl[ (3+\delta)(\epsilon+\delta) + \xi \bigr] , 
\label{rollm}
\end{equation}
and to leading order in this limit, we have just 
\begin{equation}
m^2 \approx - 3 H^2 [\epsilon+\delta] 
\label{rollmappr}
\end{equation}
and 
\begin{equation}
a(\eta) \propto {1\over\eta^{1+\epsilon}} , 
\label{aHappr}
\end{equation}
where we treat $\epsilon$ and $\delta$ as approximately constant, since their derivatives are higher order, ${\cal O}(\epsilon^2, \epsilon\delta, \delta^2)$.  

Although the equations that determine the scalar and tensor modes are second order differential equations, one of the constants of integration in each case is fixed when we impose a canonical commutations relation with its conjugate momentum.  For example, for the scalar field, 
\begin{equation}
\bigl[ \varphi(\eta,\vec x), \pi(\eta,\vec y) \bigr] 
= i\, \delta^3(\vec x-\vec y) , 
\qquad\quad
\pi = a^2\varphi' , 
\label{etcr}
\end{equation}
translates into the following condition on the modes, 
\begin{equation}
a^2 \bigl[ \varphi_k \varphi_k^{\prime *} - \varphi_k^\prime \varphi_k^* \bigr]  
= i . 
\label{wronk}
\end{equation}
If we denote the remaining constant of integration by $f_k$, then 
\begin{equation}
\varphi_k(\eta) 
= {1\over\sqrt{1-f_kf_k^*}} {\sqrt{\pi}\over 2} {H(-\eta)^{3/2}\over 1+\epsilon} 
\bigl[ H_\nu^{(2)}(k\eta) + f_k\, H_\nu^{(1)}(k\eta) \bigr] 
\label{scalarsoln}
\end{equation}
where
\begin{equation}
\nu \approx {3\over 2} + 2\epsilon + \delta 
\label{nudef}
\end{equation}
to first order in $\epsilon$ and $\delta$.

For the tensor modes, the equal-time commutator is complicated by the additional spin-structure of the fields, which allows us to construct a few more invariants than in the scalar case \cite{weinberg}\footnote{To agree with the conventions used by the Wilkinson Microwave Anisortopy Probe, or WMAP, experiment \cite{wmap}, we have not included a factor of ${1\over 2}$ in the definition of ${\cal D}_{ij,kl}$ as in \cite{weinberg}.}
\begin{equation}
\bigl[ \tau_{ij}(\eta,\vec x), a^2(\eta)\tau'_{kl}(\eta,\vec y) \bigr] 
= i\, {\cal D}_{ij,kl}(\vec x-\vec y) , 
\label{etcrT}
\end{equation}
where 
\begin{eqnarray}
{\cal D}_{ij,kl}(\vec x-\vec y) 
&\!\!\!=\!\!\!& 
\int {d^3\vec k\over(2\pi)^3}\, e^{i\vec k\cdot(\vec x-\vec y)}\, 
\label{calDdef} \\
&&
\Bigl[ 
\bigl( \delta_{ik} - (k_ik_k/k^2) \bigr)
\bigl( \delta_{jl} - (k_jk_l/k^2) \bigr)
\nonumber \\
&&
+ \bigl( \delta_{il} - (k_ik_l/k^2) \bigr)
  \bigl( \delta_{jk} - (k_jk_k/k^2) \bigr)
\nonumber \\
&&
- \bigl( \delta_{ij} - (k_ik_j/k^2) \bigr)
  \bigl( \delta_{kl} - (k_kk_l/k^2) \bigr)
\Bigr] . 
\nonumber 
\end{eqnarray}
The structure of the commutator simplifies greatly when we contract the indices of the field and its conjugate momentum, 
\begin{equation}
\bigl[ \tau_{ij}(\eta,\vec x), a^2(\eta)\tau^{\prime ij}(\eta,\vec y) \bigr] 
= 4 i\, \delta^3(\vec x-\vec y) . 
\end{equation}
Here, the factor of four simply counts the {\it two\/} physical degrees of freedom each of which appears {\it twice\/} in the matrix $\tau_{ij}$.  This relation is satisfied as long as 
\begin{equation}
a^2 \sum_{r=1}^2 \bigl[ \tau_k^{(r)} \tau_k^{(r)\prime *} 
- \tau_k^{(r)\prime} \tau_k^{(r)*} \bigr]  
= 2 i , 
\label{wronkT}
\end{equation}
which we shall assume requires that each polarization itself has a conventional normalization, 
\begin{equation}
a^2 \bigl[ \tau_k^{(r)} \tau_k^{(r)\prime *} 
- \tau_k^{(r)\prime} \tau_k^{(r)*} \bigr] = i .  
\label{wronkTr}
\end{equation}
Defining integration constants for the two polarizations, $t_k^{(r)}$, the general solution for the tensor modes is thus
\begin{eqnarray}
\tau_k^{(r)}(\eta) 
&\!\!\!=\!\!\!& 
{1\over\sqrt{1-t_k^{(r)}t_k^{(r)*}}} {\sqrt{\pi}\over 2} {H(-\eta)^{3/2}\over 1+\epsilon} 
\nonumber \\
&&\times 
\bigl[ H_n^{(2)}(k\eta) + t_k^{(r)}\, H_n^{(1)}(k\eta) \bigr] 
\label{tensorsoln}
\end{eqnarray}
where the index of the Hankel functions associated with the tensor modes is
\begin{equation}
n \approx {3\over 2} + \epsilon  
\label{tildnudef}
\end{equation}
in the slowly rolling limit.

Until now we have only constrained the form of the quantum fluctuations by appealing to general properties---the spatial flatness of the background or the canonical commutation relation, for example.  These properties allowed us to determine the mode functions up to a single remaining constant of integration in each case, $f_k$ and $t_k^{(r)}$.  Many of the studies of the potential ``trans-Planckian'' signals from inflation have introduced the effects from beyond the inflationary Hubble scale through these constants \cite{transplanck,effectivestate,schalm}.  Before examining these more exotic possibilities, we first review what assumptions underlie what are usually considered the standard predictions of inflation.  There, the basic assumption is that space-time looks flat at sufficiently small distances.  From a classical perspective, this proposition seems quite reasonable; but it fits less neatly into a quantum picture for gravity.

\subsection{The standard picture}

To fix the states completely, we must make a further guess about how nature behaves.  Usually, we make an assumption about some asymptotic property of our universe---about its infinite past or at infinitesimally short distances, for example---which then determines the constants $f_k$ and $t_k^{(r)}$.  

It is important to emphasize that even the standard picture requires some such assumption.  The standard technique for fixing the state is to choose the modes so that they match with the Minkowski vacuum modes at extremely short distances, which in this setting means $k\gg aH$ along a constant-time surface.  This procedure defines the ``Bunch-Davies vacuum'' \cite{bunch}.  Since it applies equally well to both polarizations of the tensor modes, both will be the same in this state, 
\begin{equation}
\tau_k(\eta) = \tau_k^{(1)}(\eta) = \tau_k^{(2)}(\eta) . 
\label{taugendef}
\end{equation}
Actually, whenever we assume that the symmetry between the two tensor polarizations is preserved, we shall similarly write both polarizations as $\tau_k(\eta)$, even when speaking of a more general state.

Since this state is used in most inflationary calculations, we have implicitly defined our own generic modes with respect to it.  To see this fact more clearly, let us convert for a moment from conformal time to a cosmological time, defined by 
\begin{equation}
dt = a(\eta)\, d\eta .
\label{cosmotime}
\end{equation}
Integrating this expression for the slowly rolling case, $a(\eta) \propto \eta^{-(1+\epsilon)}$, and choosing the constants of integration as well as the units of time appropriately, we can write
\begin{equation}
\eta = - {1\over{\cal H}} 
\bigl( 1 + \epsilon{\cal H} t \bigr)^{-{1\over\epsilon}} , 
\label{conftime}
\end{equation}
where ${\cal H}$ is a constant.  In the limit, $\epsilon\to 0$, 
\begin{equation}
\eta = - {e^{-{\cal H}t}\over{\cal H}} 
\biggl[ 1 + {1\over 2} \epsilon ({\cal H} t)^2 + {\cal O}(\epsilon^2) \biggr] , 
\label{conftimelim}
\end{equation}
${\cal H}$ essentially becomes  the Hubble scale $H$, which is also a constant when $\epsilon=0$.  

Let us next choose $t=0$ to be an ``initial'' time at which we define the state.  If we look at times close to this initial time, ${\cal H}t\ll 1$, then we can also expand the exponential factor $e^{-{\cal H}t}$ to obtain, 
\begin{equation}
\eta = - {1\over{\cal H}} + t + {\cal O}({\cal H} t)^2 . 
\label{conftimelimlim}
\end{equation}
Note that it is important in an inflationary background to choose a specific initial time.  The expansion of the background continually stretches the physical sizes of the modes, so what might have once been a tiny mode will, after a sufficient amount of time, no longer be so.  

At this initial time $\eta(t=0)$, the short distance limit corresponds to 
\begin{equation}
{k\over a} \gg H = {a'\over a^2} ,
\label{shortlim}
\end{equation}
which for a slowly rolling field becomes the requirement, 
\begin{equation}
- k\eta \gg 1 + \epsilon ,
\label{shortlimalt}
\end{equation}
or more simply, $k\eta\to -\infty$.  We can now expand the Hankel functions in the expression for the scalar modes, Eq.~(\ref{scalarsoln}), in this limit to discover that 
\begin{eqnarray}
\varphi_k(t\approx 0) 
&\!\!\!\!\!\approx\!\!\!\!\!& 
{1\over\sqrt{1-f_kf_k^*}} {-i\over 1+\epsilon} {H\over{\cal H}}
\nonumber \\
&&
\biggl[ 
{e^{-ikt}\over\sqrt{2k}} e^{i{k\over{\cal H}}} 
e^{i{\pi\over 2}(\nu-{3\over 2})}
+ f_k\, {e^{ikt}\over\sqrt{2k}} e^{-i{k\over{\cal H}}} 
e^{-i{\pi\over 2}(\nu-{3\over 2})} 
\biggr] . 
\nonumber \\
&&
\label{scalarsolnlim}
\end{eqnarray}
The standard flat-space vacuum state for a massless scalar field has the form 
\begin{equation}
{e^{-ikt}\over\sqrt{2k}} , 
\label{Minkmodes}
\end{equation}
so choosing the Bunch-Davies state is equivalent to setting $f_k=0$.

The calculation for the tensor modes is essentially the same, so we can conclude that $t_k^{(r)}=0$.  Therefore, the eigenmodes associated with the Bunch-Davies vacuum are
\begin{eqnarray}
\varphi_k(\eta) 
&\!\!\!=\!\!\!&  {\sqrt{\pi}\over 2(1+\epsilon)} H(-\eta)^{3/2} H_\nu^{(2)}(k\eta) 
\nonumber \\
\tau_k(\eta) 
&\!\!\!=\!\!\!& 
{\sqrt{\pi}\over 2(1+\epsilon)} H(-\eta)^{3/2} H_n^{(2)}(k\eta)  
\label{BDmodes}
\end{eqnarray}
in the slowly rolling limit.

Having thus fixed the state of the fields, we can evaluate the expected forms for the power spectra of the scalar and the tensor modes.  For the former we have
\begin{eqnarray}
P_k^\varphi(\eta) 
&\!\!\!=\!\!\!& 
{k^3\over 2\pi^2} \varphi_k(\eta)\varphi_k^*(\eta) 
\nonumber \\
&\!\!\!=\!\!\!& 
{H^2\over 8\pi} 
{(-k\eta)^3\over (1+\epsilon)^2} 
H_\nu^{(2)}(k\eta) H_\nu^{(1)}(k\eta)
\label{Pphimodes}
\end{eqnarray} 
while for the latter, 
\begin{eqnarray}
P_k^h(\eta) 
&\!\!\!=\!\!\!& 
16\pi G\, {k^3\over\pi^2} \bigl[ 
|\tau_k^{(1)}(\eta)|^2 + |\tau_k^{(2)}(\eta)|^2 \bigr] 
\nonumber \\
&\!\!\!=\!\!\!& 
16\pi G\, {2k^3\over\pi^2} \tau_k(\eta)\tau_k^*(\eta) 
\nonumber \\
&\!\!\!=\!\!\!& 
16\pi G\, {H^2\over 2\pi} {(-k\eta)^3\over (1+\epsilon)^2} 
H_n^{(2)}(k\eta) H_n^{(1)}(k\eta) . 
\label{Phmodes}
\end{eqnarray}

Inflation is essentially a mechanism for reversing the usual ordering of the rates at which physical scales and the Hubble horizon ($1/H$) grow, when compared with a radiation or matter dominated universe.  By the end of the inflationary era, all of the modes responsible for the structures we observe at large scales should have been stretched well outside the horizon.  For these modes, we must consider the opposite limit to that used to define the states, 
\begin{equation}
{k\over a} \ll H = {a'\over a^2} 
\qquad\hbox{or}\qquad 
k \ll {a'\over a} . 
\label{stretched}
\end{equation}
Again, in the slowly rolling regime this condition translates into 
\begin{equation}
- k\eta \ll 1+\epsilon , 
\label{SLstretch}
\end{equation}
or just, $-k\eta \ll 1$.

Expanding the Bunch-Davies power spectra for this limit at last produces the standard predictions for the scalar and tensor perturbations for inflation.  Thus, for the scalar modes we have 
\begin{eqnarray}
P_k^\varphi(\eta)
&\!\!\!=\!\!\!& 
{H^2\over 4\pi^2} (-k\eta)^{- 4\epsilon - 2\delta} 
\nonumber \\
&&\times
\bigl[ 1 - 2\epsilon + 2 (2\epsilon + \delta) [ 2 - \gamma - \ln 2 ]
+ \cdots \bigr] ;
\qquad
\label{slowPks}
\end{eqnarray}
and similarly, we should have a faint background of primordial gravity waves too, 
\begin{eqnarray}
P_k^h(\eta)
&\!\!\!=\!\!\!& 
16\pi G\, {H^2\over \pi^2} (-k\eta)^{-2\epsilon} 
\nonumber \\
&&\times
\bigl[ 1 + 2\epsilon [ 1 - \gamma - \ln 2 ]
+ \cdots \bigr] . 
\label{slowPkt}
\end{eqnarray}
Notice that both of these power spectra are nearly flat---their dependence on $k$ is quite weak.  The leading departure of these patterns form perfect Gaussian noise, at least at the level of the two-point function, is often described by a tilt parameter, defined with slightly different conventions for the scalar and tensor cases,
\begin{eqnarray}
n_S - 1 &\!\!\!\equiv\!\!\!& {d\ln P_k^\varphi\over d\ln k} 
\nonumber \\ 
n_T &\!\!\!\equiv\!\!\!& {d\ln P_k^h\over d\ln k} . 
\label{tilts}
\end{eqnarray}
Evaluating these for a period of slowly rolling inflation yields 
\begin{eqnarray}
n_S &\!\!\!=\!\!\!& 1 - 4\epsilon - 2\delta 
\nonumber \\ 
n_T &\!\!\!=\!\!\!& - 2\epsilon . 
\label{tiltsSR}
\end{eqnarray}

Experimentally, the power spectrum is usually measured using a $k_0$ as a reference point.\footnote{In the WMAP experiment, for example, its value is $k_0 = 2\ {\rm Gpc}^{-1}$ \cite{wmap}.}  So we can make the following very simple rescalings, 
\begin{eqnarray}
P_k^\varphi(\eta)
&\!\!\!\!\!=\!\!\!\!\!& 
\Delta^2_\varphi(k_0\eta) \biggl( {k\over k_0} \biggr)^{- 4\epsilon - 2\delta} 
\nonumber \\
P_k^h(\eta)
&\!\!\!\!\!=\!\!\!\!\!& 
\Delta^2_h(k_0\eta) \biggl( {k\over k_0} \biggr)^{-2\epsilon} , 
\label{powersk0}
\end{eqnarray}
where of course, 
\begin{eqnarray}
\Delta^2_\varphi(k_0\eta)
&\!\!\!\!\!=\!\!\!\!\!& 
{H^2\over 4\pi^2} (-k_0\eta)^{- 4\epsilon - 2\delta} 
\nonumber \\
&&\times
\bigl[ 1 - 2\epsilon + 2 (2\epsilon + \delta) [ 2 - \gamma - \ln 2 ]
+ \cdots \bigr] .
\qquad
\label{slowPksk0}
\end{eqnarray}
and 
\begin{equation}
\Delta^2_h(k_0\eta) = 16\pi G\, {H^2\over \pi^2} (-k_0\eta)^{-2\epsilon} 
\bigl[ 1 + 2\epsilon [ 1 - \gamma - \ln 2 ]
+ \cdots \bigr] . 
\label{slowPktk0}
\end{equation}

\subsection{The scalar to tensor ratio}

When we expanded the action, we rescaled the fields so that they would assume a standard normalization for their kinetic terms.  This procedure defined the scalar and tensor fields $\varphi$ and $\tau_{ij}$, and it allowed us to appeal to the canonical commutation relations to set the normalization of their eigenmodes.  However, when we wish to compare the predictions of inflation to what is actually measured, we should return to the appropriately normalized fields.  What is most important is how the perturbations affect the gravitational evolution of the universe, so we should rescale the fields to restore the normalization of the original gravitational action from which they arose.  For the tensor fields, we need only switch back from $\tau_{ij}$ to $h_{ij}$, which was already done when we wrote the power spectrum $P_k^h$ rather than $P_k^\tau$.

The scalar field is slightly more subtle, being a linear combination of a piece of the metric and the inflaton.  From its definition in Eq.~(\ref{varphidef}), we can see that the metric part, $\Psi$, is weighted with a factor of $\phi'_0/aH$; so we can put the scalar modes on the same footing as the tensor modes by rescaling by this factor, thereby defining a new field $\zeta$, 
\begin{equation}
\zeta \equiv {aH\over\phi'_0} \varphi 
= \Psi + {aH\over\phi'_0} \delta\varphi . 
\label{zetadef}
\end{equation}
$\zeta$ is still invariant under small coordinate transformations.  We can thus obtain the power spectrum for $\zeta$ very easily,
\begin{equation}
P_k^\zeta(\eta)
= \biggl( {aH\over\phi'_0} \biggr)^2 P_k^\varphi(\eta)
= \Delta^2_\zeta(k_0\eta) \biggl( {k\over k_0} \biggr)^{- 4\epsilon - 2\delta} , 
\label{powerzeta}
\end{equation}
where 
\begin{eqnarray}
\Delta^2_\zeta(k_0\eta)
&\!\!\!=\!\!\!& 
\biggl( {aH\over\phi'_0} \biggr)^2 P_\varphi(k_0\eta)
\nonumber \\
&\!\!\!=\!\!\!& 
{4\pi G\over\epsilon}\, {H^2\over 4\pi^2} (-k_0\eta)^{- 4\epsilon - 2\delta} 
\nonumber \\
&&\times
\bigl[ 1 - 2\epsilon + 2 (2\epsilon + \delta) [ 2 - \gamma - \ln 2 ]
+ \cdots \bigr] .
\qquad
\label{powerzetak0}
\end{eqnarray}
Note that here we have used the relation
\begin{equation}
aH' = - 4\pi G\, (\phi'_0)^2 
\quad\Rightarrow\quad 
\biggl( {aH\over\phi'_0} \biggr)^2 = {4\pi G\over\epsilon} , 
\label{relation}
\end{equation}
which follows from the unperturbed Einstein equation---its $\eta\eta$ component, to be exact---along with the equation of motion of the zero mode of the scalar field, $\phi_0$.

Finally, we define the {\it scalar to tensor ratio\/} by taking quotient of the amplitudes of power spectra, 
\begin{equation}
r \equiv {\Delta^2_h(k_0\eta)\over \Delta^2_\zeta(k_0\eta)} , 
= 16 \epsilon , 
\label{ratiodef}
\end{equation}
keeping only the leading contribution in the slowly rolling limit.  Because of how we have defined the states for the scalar and tensor fluctuations, this ratio has a fixed proportionality to $\epsilon$, which is sometimes expressed by the following relation with the tilt of the gravity wave spectrum,
\begin{equation}
n_T = - {r\over 8} . 
\label{consist}
\end{equation}

\section{A simple illustration in de Sitter space}
\label{deSitter}

In the last section, we showed the detailed assumptions that lie behind the standard prediction from inflation for the scalar to tensor ratio.  A central assumption is that the states for both the scalar and tensor modes can be reasonably described by a Bunch-Davies state or, equivalently, that nature at very short distances during the inflationary era should resemble a Minkowski vacuum.  By a ``short'' distance, we mean one where the curvature of the space-time is not much apparent.  For example, for a field this condition refers to Fourier modes whose wavelength---{\it at a particular time\/}---are much smaller than the size of the Hubble horizon, $1/H$, 
\begin{equation}
{k\over a(\eta)} \ll H(\eta) . 
\label{shortdist}
\end{equation}
Aside from some very general and mild bounds, there are few direct experimental constraints on the value of $H$.  It might be anywhere between about $10^{14}$ GeV, beyond which the amplitude of the scalar perturbations would be larger than what is observed\footnote{Here we have implicitly used the constraint on $r<0.20$ \cite{wmap} to constrain $\epsilon$.  This bound might change if the simple relation between $r$ and $\epsilon$ is broken.}, down to about a TeV, where the mechanism driving inflation ought to have already appeared in accelerator experiments.  Note that at the upper end of this range, the Hubble scale is no longer an infinitesimal fraction of the Planck scale $M_{\rm pl}$, defined by 
\begin{equation}
 M_{\rm pl} = {1\over\sqrt{8\pi G}} \sim 2.4 \times 10^{18}\ {\rm GeV} . 
\label{planckdef}
\end{equation}
So in taking the ``short-distance'' limit, we are appealing to the behavior at scales approaching a Planck length.  And since inflation essentially relies on a {\it quantum\/} mechanism for generating the primordial perturbations, it might be a little worrisome to define the state near a region where we lack an adequate description for the quantum behavior of gravity.

The condition on the modes is also a time-dependent one.  The evolution of the modes is an essential ingredient of inflation, since it allows small, sub-horizon modes to be stretched to much larger scales where they are effectively frozen into the background, at least until some much later epoch of the universe.  The expansion implies that for a long enough period of inflation, every mode whose influence has been implicitly observed in the patterns of the universe would have once had a wavelength smaller than the Planck length.

A general inflationary background, while still invariant under spatial translations and rotations, has fewer symmetries than Minkowski space.  To model some of the peculiar behavior that could occur at the Planckian regime, here we shall examine operators that break the coordinate invariance of the action while still maintaining these spatial symmetries \cite{planck}.  To keep the analysis relatively simple, in this section we study the idealized example of a de Sitter background, where the scale factor becomes
\begin{equation}
a(\eta) = - {1\over H\eta} .
\label{dSscale}
\end{equation}
with a {\it constant\/} Hubble scale.  Although this background actually does have the same number of space-time symmetries as flat space---though still lacking a globally time-like symmetry---we shall nevertheless look at the effect of one of the symmetry-breaking operators, as a useful illustration.

To begin, since when we originally expanded the action we only retained terms that are quadratic in the fields $\varphi$ and $\tau_{ij}$, we shall similarly limit ourselves to symmetry-breaking operators that are quadratic in these fields.  Very generally, the dimension-four operators that preserve spatial translations and rotations can be classified according to the number of time or space derivatives that they contain.  Since the Hubble scale in de Sitter space is a constant, an operator that reduces to the form $H^2\varphi^2$ acts as a mass term and we shall neglect it.  We are left with two other possibilities at this order,
\begin{equation}
{H\over a(\eta)} \varphi \bigl( - \vec\nabla\cdot\vec\nabla \bigr)^{1/2} \varphi 
\quad\hbox{and}\quad
{1\over a^2(\eta)} \vec\nabla\varphi \cdot \vec\nabla\varphi . 
\label{dimfour}
\end{equation}
Here we have introduced a non-analytical derivative operator $\bigl( - \vec\nabla\cdot\vec\nabla \bigr)^{1/2}$.  Although it might at first seem a little peculiar, its role is simply to extract the magnitude of the spatial momentum of a mode.  For example,
\begin{equation}
\bigl( - \vec\nabla\cdot\vec\nabla \bigr)^{1/2} \varphi(\eta,\vec x) 
= \int {d^3\vec k\over (2\pi)^3}\, |\, \vec k\, | \, 
\bigl[ \varphi_k e^{i\vec k\cdot\vec x} a_{\vec k} 
+ \varphi_k^* e^{-i\vec k\cdot\vec x} a_{\vec k}^\dagger \bigr] . 
\label{dhalfonphi}
\end{equation}
We shall use it since it produces exactly the same qualitative effects as a large class of higher order operators.

The first operator, which contains a single power of the derivative in the form $\bigl( - \vec\nabla\cdot\vec\nabla \bigr)^{1/2}$, has a seemingly innocuous effect on the power spectrum of its respective field.  Its correction---as we shall see---to the standard inflationary prediction has no significant dependence on the momentum.  However, the interesting feature here is that the {\it combined\/} effect of such an operator for both the scalar and the tensor fields can alter the ``consistency relation'' away from the $r=-8n_T$ that is usually expected for inflation.

To see this effect, let us consider the following interaction Hamiltonian, 
\begin{eqnarray}
H_I(\eta) 
&\!\!\!\!\!=\!\!\!\!\!& c_S\, a^3 H \int d^3\vec x\, 
\varphi \bigl( - \vec\nabla\cdot\vec\nabla \bigr)^{1/2} \varphi
\nonumber \\
&&
+\ c_T\, a^3 H \int d^3\vec x\, 
\tau_{ij} \bigl( - \vec\nabla\cdot\vec\nabla \bigr)^{1/2} \tau^{ij} , 
\label{Hinteasy}
\end{eqnarray}
which in de Sitter space reduces to 
\begin{eqnarray}
H_I(\eta) 
&\!\!\!\!\!=\!\!\!\!\!& - {c_S\over H^2\eta^3} \int d^3\vec x\, 
\varphi \bigl( - \vec\nabla\cdot\vec\nabla \bigr)^{1/2} \varphi
\nonumber \\
&&
-\ {c_T\over H^2\eta^3} \int d^3\vec x\, 
\tau_{ij} \bigl( - \vec\nabla\cdot\vec\nabla \bigr)^{1/2} \tau^{ij} . 
\label{HinteasydS}
\end{eqnarray}
Treating their role in the evolution of the power spectra perturbatively, these operators produce corrections to the power spectrum.  For the scalar field we have
\begin{eqnarray}
&&\!\!\!\!\!\!\!\!\!\!\!\!\!\!\!\!\!\!\!\!\!\!\!
\langle 0(\eta) | \varphi(\eta,\vec x) \varphi(\eta,\vec y) |0(\eta)\rangle 
\nonumber \\
&\!\!\!\!\!=\!\!\!\!\!&
\langle 0 | \varphi(\eta,\vec x) \varphi(\eta,\vec y) |0\rangle 
\nonumber \\
&& 
- i \int_{\eta_0}^\eta d\eta'\, \langle 0(\eta) | \bigl[ 
\varphi(\eta,\vec x) \varphi(\eta,\vec y), H_I(\eta') \bigr] |0(\eta)\rangle 
\nonumber \\
&&
+ \cdots , 
\label{dScorrect}
\end{eqnarray}
and we find to first order in $c_S$ that the power spectrum of the scalar modes becomes
\begin{equation}
P_k^\varphi(\eta) = {H^2\over 4\pi^2} \bigl[ 1 + \pi c_S + \cdots \bigr] , 
\label{PvarphidS}
\end{equation}
where the neglected terms are of the order ${\cal O}\bigl( (k\eta)^2, (k\eta_0)^{-1}\bigr)$.

In integrating the cumulative effect of the interaction Hamiltonian, we introduced an initial time $\eta_0$.  For now, we could just view it as a regulator, but in the next section we shall find that it has a more important role, as the earliest time for the applicability of an effective theory treatment of symmetry-breaking operators.  To be applicable to the modes responsible for the actual structures that we see at large scales in the universe, this time should be sufficiently early that the wavelength of any of these modes was smaller than the Hubble horizon at that time, 
\begin{equation}
{k\over a(\eta_0)} \gg H 
\qquad\Rightarrow\qquad 
-k\eta_0 \gg 1 .
\label{eta0def}
\end{equation}
Therefore, the appropriate limit to take is $-k\eta_0\to\infty$, at least when it is possible, and $-k\eta\to 0$, as was explained earlier.  In these limits the only effect on the power spectrum is a constant one, as mentioned above.  Similarly, for the tensor modes we find 
\begin{equation}
P_k^h(\eta) = 16\pi G\, {H^2\over \pi^2} \bigl[ 1 + 2\pi c_T + \cdots \bigr] ; 
\label{PtdS}
\end{equation}
the additional factor of two comes from the two polarizations of the gravity waves.

Such effects are usually regarded as essentially unobservable, since the presence of such an operator could not be distinguished from a standard inflationary universe with a different Hubble scale, $\tilde H = \sqrt{1+\pi c_S} H$, for example, since there is no $k$-dependence that is not explicitly suppressed.  However, such operators might not be quite so innocuous as is often stated.  Combining both the scalar and the tensor modes, we have nominally
\begin{equation}
r \to r \approx {1+2\pi c_T\over 1+\pi c_S} r_0, 
\label{shiftedr}
\end{equation}
where $r_0$ is the standard value for the scalar to tensor ratio---$r_0= 16\epsilon$ in a slowly rolling universe, for example.  Even for relatively mild values of $c_S$ and $c_T$, the previous relation between tilt of the power spectrum of the gravity waves $n_T$ and the ratio $r$ no longer needs to hold.  Still more interestingly, for $c_S<0$ (while assuming $1+\pi c_S > 0$) and $c_T$ we can have an enhancement of the gravity wave signal relative to the scalar one.  Here we have been assuming that $c_S$ and $c_T$ are small enough for a perturbative treatment, but we could relax this requirement, summing over multiple insertions of the interaction Hamiltonian, to obtain an essentially arbitrary value of $r$, with no especial relation to any of the slow-roll parameters.

This section has been intended as an illustration highlighting the effect of a particular symmetry-breaking operator in a background where the form of the mode functions is quite simple.  In de Sitter space, both the scalar and the tensor modes reduce to the same form, 
\begin{equation}
\varphi_k(\eta) = \tau_k(\eta) 
= {H\over\sqrt{2k^3}} (i-k\eta) e^{-ik\eta} . 
\label{dSmodes}
\end{equation}
However, in the process we have left the slowly rolling regime, and some of the quantities---such as $P_k^\zeta(\eta)$---actually diverge while others---such as $r$---vanish as we take $\epsilon\to 0$.  Therefore, the next section is devoted to a look at these symmetry-breaking operators in a more realistic environment.

\section{Returning to the slowly rolling regime}
\label{slow}

Some of what we derive in this section reproduces the picture that we just have seen, except that here we examine a slowly rolling epoch of inflation for which the scalar to tensor ratio is a sensible, non-vanishing quantity.  However, we also encounter some less familiar phenomena too, which produce scale ($k$) dependent modulations in the power spectra.  These oscillations can affect the ratio of their amplitudes in interesting ways.  The ratio can be seemingly enhanced or diminished by fortuitous choices of $k_0$, the scale chosen to define the amplitudes of the power spectra.

We start by writing the symmetry-breaking operators in a slightly more general form, which emphasizes the special role of the time coordinate in a uniformly expanding background.  By defining a unit vector pointing in the positive-time direction,
\begin{equation}
n_\mu = \bigl(a(\eta), 0, 0, 0\bigr) , 
\label{ndef}
\end{equation}
we can define an induced metric on the surfaces orthogonal to this direction,
\begin{equation}
h_{\mu\nu} = g_{\mu\nu} - n_\mu n_\nu ; 
\label{induceddef}
\end{equation}
in this case, they are flat surfaces, at constant values of the conformal time, 
\begin{equation}
h_{\mu\nu}\, dx^\mu dx^\nu = - a^2(\eta)\, d\vec x\cdot d\vec x . 
\label{induceddS}
\end{equation}
Since they are flat, their induced curvature vanishes.  There is a second curvature tensor associated with how these surfaces are embedded in the full space-time.  This {\it extrinsic curvature\/} is defined by 
\begin{equation}
K_{\mu\nu} = h_\mu^{\ \lambda}\nabla_\lambda n_\nu
\label{Kdef}
\end{equation}
and in the inflationary background it reduces to 
\begin{equation}
K_{\mu\nu}\, dx^\mu dx^\nu = - a^2 H\, d\vec x\cdot d\vec x . 
\label{KdS}
\end{equation}
Its trace is proportional to the Hubble scale, $K = g^{\mu\nu}K_{\mu\nu} = 3H$. 

We can also generalize the spatial derivative operator by defining an operator ${\cal D}$ by
\begin{equation}
{\cal D} \equiv \bigl( h^{\mu\nu} \nabla_\mu \nabla_\nu 
- K n^\mu \nabla_\mu \bigr)^{1/2} ; 
\label{calDdefS}
\end{equation}
it reduces to the previous form, $a^{-1}(-\vec\nabla\cdot\vec\nabla)^{1/2}$, in an isotropically expanding background.

With these objects, $h_{\mu\nu}$, $K_{\mu\nu}$ and ${\cal D}$, we now construct the general, {\it leading\/} symmetry-breaking operators for the scalar field,
\begin{equation}
{\cal L}_S = {c_S^{p,q}\over M^{p+q-2}} \biggl( {K\over 3} \biggr)^p 
\varphi {\cal D}^q \varphi , 
\label{genops}
\end{equation}
and analogously 
\begin{equation}
{\cal L}_T = {c_T^{p,q}\over M^{p+q-2}} \biggl( {K\over 3} \biggr)^p 
\tau_{ij} {\cal D}^q \tau^{ij} , 
\label{genopt}
\end{equation}
for the tensor fluctuations.  Here, $p$ and $q$ are integers.  $M$ corresponds to an energy scale associated with a short-distance breakdown of the usual space-time symmetries; in this setting, at distances smaller than $1/M$ we no longer assume that space-time is locally flat.  Alternatively, $M$ could also be the energy scale of some new dynamics, such as an excited field that couples to either $\varphi$ or $\tau_{ij}$.  

We are still implicitly working in a slowly rolling inflationary universe, so derivatives of $H$ are small compared with powers of $H$.  We have also written the general operators with all of the spatial derivatives acting on only one of the fields; but for the two-point function, which has only a single external momentum $k$, this hardly matters since the same momentum is running through each vertex.  In the isotropically expanding background, these operators reduce to 
\begin{equation}
{\cal L}_S = {c_S^{p,q}\over M^{p+q-2}} {H^p\over a^q} 
\varphi \bigl( - \vec\nabla\cdot\vec\nabla \bigr)^{q/2} \varphi , 
\label{genopseg}
\end{equation}
and analogously 
\begin{equation}
{\cal L}_T = {c_T^{p,q}\over M^{p+q-2}} {H^p\over a^q} 
\tau_{ij} \bigl( - \vec\nabla\cdot\vec\nabla \bigr)^{q/2} \tau^{ij} . 
\label{genopteg}
\end{equation}

To avoid defining our state in an era when the modes in which we are interested---those producing the patterns in the microwave background radiation---have physical wavelengths smaller than a Planck length, we shall start the time-evolution of the two-point functions at a time $\eta_0$ when the wavelengths of these modes are still larger than this length.  In the standard Bunch-Davies state, these operators produce the following corrections to the power spectra of $\zeta$ and $h_{ij}$,\footnote{Throughout this section we are retaining at most only the first corrections in the slowly rolling limit.} 
\begin{eqnarray}
\delta P_k^\zeta(\eta) 
&\!\!\!\!\!=\!\!\!\!& 
{4\pi G\over\epsilon} {c_S^{p,q}\over 16} (1-4\epsilon) {k^q\over M^{p+q-2}} 
\nonumber \\
&&
H^2(\eta) (-k\eta)^3
\int_{\eta_0}^\eta d\eta'\, (-\eta')^3 a^{4-q}(\eta') H^{p+2}(\eta') 
\nonumber \\
&&
i \bigl\{ \bigl[ H_\nu^{(2)}(k\eta) H_\nu^{(1)}(k\eta') \bigr]^2
- \bigl[ H_\nu^{(1)}(k\eta) H_\nu^{(2)}(k\eta') \bigr]^2 \bigl\}
\nonumber \\
&&
\label{gendPscalar}
\end{eqnarray}
and 
\begin{eqnarray}
\delta P_k^h(\eta) 
&\!\!\!\!\!=\!\!\!\!& 
16\pi G {c_T^{p,q}\over 2} (1-4\epsilon) {k^q\over M^{p+q-2}} 
\nonumber \\
&&
H^2(\eta) (-k\eta)^3
\int_{\eta_0}^\eta d\eta'\, (-\eta)^3 a^{4-q}(\eta') H^{p+2}(\eta') 
\nonumber \\
&&
i \bigl\{ \bigl[ H_n^{(2)}(k\eta) H_n^{(1)}(k\eta') \bigr]^2
- \bigl[ H_n^{(1)}(k\eta) H_n^{(2)}(k\eta') \bigr]^2 \bigl\} . 
\nonumber \\
&&
\label{gendPtensor}
\end{eqnarray}
Both integrals are of the same basic form, the only difference being in the indices of the Hankel functions.  Their integrands are also completely finite throughout the range, 
\begin{equation}
-\infty < \eta_0 \le \eta' \le \eta < 0 ,
\label{etaprrange}
\end{equation}
and the only possibilities for divergences occur at the limits, $\eta\to 0$ and $\eta_0\to -\infty$.  

To study these potential divergences more carefully, let us introduce a mass scale ${\cal H}$ so that we can write the scale factor in the form, 
\begin{equation}
a(\eta) = {1\over (-{\cal H}\eta)^{1+\epsilon}} . 
\label{ainSR}
\end{equation}
Since it is a constant, ${\cal H}$ is not quite equal to the Hubble scale, 
\begin{equation}
H(\eta) = (1+\epsilon) {\cal H}^{1+\epsilon} (-\eta)^\epsilon . 
\label{HinSR}
\end{equation}
Finally, since a large or small $\eta$ has a physical meaning only when compared with the wave number $k$, let us define the following {\it positive\/} dimensionless variables, 
\begin{equation}
x \equiv -k\eta,
\quad 
x' \equiv -k\eta'
\quad\hbox{and}\quad 
x_0 \equiv -k\eta_0 , 
\label{xdefs}
\end{equation}
in terms of which the leading correction to the scalar power spectrum becomes, 
\begin{eqnarray}
\delta P_k^\zeta(x) 
&\!\!\!\!\!=\!\!\!\!& 
{4\pi G\over\epsilon} {c_S^{p,q}\over 16} (1 + p\epsilon) 
{{\cal H}^{(p+q)(1+\epsilon)}\over M^{p+q-2}} 
k^{-\epsilon(p+q)} 
\nonumber \\
&&
x^{3+2\epsilon} 
\int_x^{x_0} dx'\, (x')^{-1 + q + \epsilon(-2+p+q)} 
\nonumber \\
&&
i \bigl\{ \bigl[ H_\nu^{(1)}(x) H_\nu^{(2)}(x') \bigr]^2
- \bigl[ H_\nu^{(2)}(x) H_\nu^{(1)}(x') \bigr]^2 \bigl\} , 
\nonumber \\
&&
\label{gendPsxs}
\end{eqnarray}
with an analogous result for $P_k^h(x)$.

A mode stretched far outside the Hubble horizon by the end of inflation is one for which $x = - k\eta \to 0$, so we can expand the Hankel functions in this limit, yielding 
\begin{eqnarray}
\delta P_k^\zeta(x) 
&\!\!\!=\!\!\!& 
- {4\pi G\over\epsilon} {{\cal H}^2\over 4\pi^2} 
2^{2\nu} \Gamma^2(\nu) (1 + p\epsilon) 
\nonumber \\
&&
c_S^{p,q}
\biggl( {{\cal H}\over M} \biggr)^{p+q-2} 
\biggl( {{\cal H}\over k} \biggr)^{\epsilon(p+q)} 
x^{3 - 2\nu + 2\epsilon} 
\nonumber \\
&&
\int_x^{x_0} dx'\, (x')^{-1 + q + \epsilon(-2+p+q)} J_\nu(x') Y_\nu(x') ,  
\qquad
\label{gendPsxsmall}
\end{eqnarray}
along with terms that explicitly vanish at $x=0$.  Next, expanding the integrand in the limit where $x'\to x\to 0$, the leading behavior is 
\begin{eqnarray}
&&\!\!\!\!\!\!\!\!\!\! 
(x')^{-1 + q + \epsilon(-2+p+q)} J_\nu(x') Y_\nu(x') 
\nonumber \\
&& 
= - {1\over \pi\nu} (x')^{-1 + q + \epsilon(-2+p+q)} 
\bigl[ 1 + {\cal O}(x^{\prime 2}) \bigr] . 
\label{aympx0}
\end{eqnarray}
Its integral remains finite as long as the operator contains at least one power of the spatial derivative, $q\ge 1$.  As in de Sitter space, the $q=0$ case more or less resembles a mass term, since the Hubble parameter is still approximately constant.  Therefore, we shall always restrict to the $q\ge 1$ case, where we can therefore safely let these modes expand until they are well outside of the horizon.

Next consider how the various corrections to the power spectrum from these symmetry-breaking operators scale with $x_0$.  The parameter $x_0$ counts the number of wavelengths of a given mode that fit within the horizon at $\eta_0$.  Since inflation is intended to provide a causal explanation for the correlated structures seen throughout the universe, we always have $x_0>1$.  Modes that are well within the horizon at $\eta_0$, will correspondingly satisfy $x_0\gg 1$, thus we shall usually consider the behavior of the corrections as $x_0\to\infty$.

In the limit $x'\sim x_0\to\infty$, the leading parts of the integrand contribute, 
\begin{eqnarray}
\delta P_k^\zeta(x_0) 
&\!\!\!=\!\!\!& 
- {4\pi G\over\epsilon} {{\cal H}^2\over 4\pi^2} (1+p\epsilon) 
\bigl[ 1 + (2\nu-3)(2-\gamma-\ln 2) \bigr]
\nonumber \\
&& 
c_S^{p,q} \biggl( {{\cal H}\over M} \biggr)^{p+q-2} 
\biggl( {{\cal H}\over k} \biggr)^{\epsilon(p+q)} 
x^{3-2\nu+2\epsilon} 
\nonumber \\
&& 
\int^{x_0} dx'\, 
x^{\prime - 2 + q + \epsilon(-2+p+q)} 
\nonumber \\
&&\qquad\times
\bigl\{ \sin(2x') - {\textstyle{1\over 2}}\pi \bigl( 2\nu - 3 \bigr) \cos(2x') \bigr\} , 
\label{Pzetax0}
\end{eqnarray}
for the scalar modes and 
\begin{eqnarray}
\delta P_k^h(x_0) 
&\!\!\!=\!\!\!& 
- 16\pi G {2{\cal H}^2\over\pi^2} (1+p\epsilon) 
\bigl[ 1 + (2n-3)(2-\gamma-\ln 2) \bigr]
\nonumber \\
&& 
c_T^{p,q} \biggl( {{\cal H}\over M} \biggr)^{p+q-2} 
\biggl( {{\cal H}\over k} \biggr)^{\epsilon(p+q)} 
x^{3-2n+2\epsilon} 
\nonumber \\
&& 
\int^{x_0} dx'\, 
x^{\prime - 2 + q + \epsilon(-2+p+q)} 
\nonumber \\
&&\qquad\times
\bigl\{ \sin(2x') - {\textstyle{1\over 2}}\pi \bigl( 2n - 3 \bigr) \cos(2x') \bigr\} , 
\label{Phx0}
\end{eqnarray}
for the tensor modes.  In the slowly rolling limit, both $\nu$ and $n$ are nearly equal to $3\over 2$,
\begin{equation}
\nu - {\textstyle{3\over 2}} = 2\epsilon + \delta, 
\qquad
n - {\textstyle{3\over 2}} = \epsilon .
\label{nandnu}
\end{equation}
If we ignore corrections which are suppressed in this limit, both of the integrands have the form $(x')^{-2+q}\sin(2x')$ which can be readily evaluated.  The leading corrections for the scalar and tensor power spectra are then
\begin{equation}
\delta P_k^\zeta(x_0) 
= {4\pi G\over\epsilon} {H^2\over 8\pi^2}
x^{- 2\epsilon - 2\delta} 
c_S^{p,q} \biggl( {H\over M} \biggr)^{p+q-2} 
x_0^{q-2}\, \cos(2x_0) 
\label{Pzetaflatroll}
\end{equation}
and
\begin{equation}
\delta P_k^h(x_0) 
= 16\pi G {H^2\over\pi^2} 
c_T^{p,q} \biggl( {H\over M} \biggr)^{p+q-2} 
x_0^{q-2}\, \cos(2x_0) . 
\label{Phflatroll}
\end{equation}
Since we are ignoring ${\cal O}(\epsilon, \delta)$ corrections in these equation, we have replaced the (constant) ${\cal H}$ with the (weakly time-dependent) $H$.

Although $x_0$ has a simple enough interpretation, these corrections become a little more transparent if we introduce a $k_\star$ associated with a mode whose wavelength is exactly equal to $1/M$ at the start of the evolution, $\eta_0$, 
\begin{equation}
{k_\star\over a(\eta_0)} = M . 
\label{kstardef}
\end{equation}
To avoid modes that whose effects lie outside the applicability of the perturbative description that we have been using, we should limit ourselves to those with $k<k_\star$, which for a slowly rolling universe means 
\begin{equation}
\eta_0 = - {1\over{\cal H}} 
\biggl( {M\over k_\star} \biggr)^{{1\over 1+\epsilon}} . 
\label{kstardefSR}
\end{equation}
or, more simply, 
\begin{equation}
x_0 = {M\over H} {k\over k_\star} , 
\label{kstardefdS}
\end{equation}
neglecting ${\cal O}(\epsilon)$ corrections.  In terms of this maximal wavenumber, the corrections to the power spectra become 
\begin{eqnarray}
\delta P_k^\zeta(x_0) 
&\!\!\!=\!\!\!& 
{4\pi G\over\epsilon} {H^2\over 8\pi^2}
x^{- 2\epsilon - 2\delta} 
\nonumber \\
&&
c_S^{p,q} \biggl( {H\over M} \biggr)^p 
\biggl( {k\over k_\star} \biggr)^{q-2}
\, \cos\biggl( 2 {M\over{\cal H}}{k\over k_\star} \biggr) 
\quad
\label{Pzetaflatrollks}
\end{eqnarray}
and
\begin{eqnarray}
\delta P_k^h(x_0) 
&\!\!\!=\!\!\!& 
16\pi G {H^2\over\pi^2} 
\nonumber \\
&&
c_T^{p,q} \biggl( {H\over M} \biggr)^p 
\biggl( {k\over k_\star} \biggr)^{q-2}
\, \cos\biggl( 2 {M\over{\cal H}}{k\over k_\star} \biggr) . 
\quad
\label{Phflatrollks}
\end{eqnarray}
In deriving these correction, we have been assuming that the original integral over $d\eta'$ diverged as we allowed the initial time to be taken arbitrarily early, $\eta_0\to -\infty$, or equivalently $k/k_\star\to\infty$.  Therefore, these expressions do not apply to the $q<2$ case.

In the next section we shall discuss specific cases of these corrections to the power spectrum and how they would affect the scalar to tensor ratio.  The dimension of a given symmetry-breaking operator that produces these corrections is
\begin{equation}
d = p + q + 2 ,
\label{dimofop}
\end{equation}
so we can equivalently label the operators at a given order by $d$ and $q=1,\ldots d-2$, omitting the $q=0$ case since it behaves like a mass term.  Thus, in summary, the leading corrections to the scalar power spectrum are, neglecting small, slowly-rolling corrections 
\begin{widetext}
\begin{equation}
\delta P^\zeta_k 
= {4\pi G\over\epsilon} {H^2\over 4\pi^2} 
\biggl( {H\over M} \biggr)^{d-4} 
\times\cases{
\pi\, c_S^{p,1} + \cdots 
&$q=1$ \cr
{\displaystyle 
c_S^{p,2} 
\biggl[ 3 + \cos\biggl( 2 {M\over H} {k\over k_\star} \biggr) \biggr] 
+ \cdots }
&$q=2$\cr
{\displaystyle 
{1\over 2} c_S^{p,q}\, 
\biggl( {M\over H} {k\over k_\star} \biggr)^{q-2}
\cos\biggl( 2 {M\over H}{k\over k_\star} \biggr) 
+ \cdots }
&$q>2$\cr} , 
\label{Pzetacases}
\end{equation}
while the corrections to the power spectrum of the gravity waves are 
\begin{equation}
\delta P^h_k 
= 16\pi G {H^2\over\pi^2} \biggl( {H\over M} \biggr)^{d-4} 
\times
\cases{
2\pi\, c_T^{p,1} + \cdots 
&$q=1$ \cr
{\displaystyle 
2\, c_T^{p,2} 
\biggl[ 3 + \cos\biggl( 2 {M\over H} {k\over k_\star} \biggr) \biggr] 
+ \cdots }
&$q=2$\cr
{\displaystyle 
c_T^{p,q}\, 
\biggl( {M\over H} {k\over k_\star} \biggr)^{q-2}
\cos\biggl( 2 {M\over H}{k\over k_\star} \biggr) 
+ \cdots }
&$q>2$\cr} . 
\label{Phcases}
\end{equation}
\end{widetext}

\section{Discussion}
\label{discuss}

Any evidence of primordial gravity waves would likely be seen by many as a confirmation that the universe underwent an inflationary expansion.  However, while such an observation could effectively rule out some ideas for what occurred in the early universe, that is not quite the same as claiming that it has confirmed others.  At best, we can only say that a theory is still {\it consistent\/} with observations.  As one example, even in a universe where inflation did occur, it is still possible that the observable gravity wave background could have nothing to do with the tensor modes generated during inflation.  A phase transition after inflation ended, for instance, could easily overwhelm the inflationary gravity-wave background \cite{krauss}.  Even if our own universe is not as perverse as this example, we would still like to be able to state precisely what we can infer if future experiments were to observe a nonvanishing scalar to tensor ratio.  What is the standard, or minimal, prediction of inflation?  Can inflation itself naturally produce departures from this minimal prediction?  And where ought we to look next to distinguish among the possible mechanisms for producing these departures?

One standard prediction for a slowly rolling model with a single inflaton field is that the scalar to tensor ratio should be
\begin{equation}
r = 16\epsilon = -16 {H'\over aH^2} . 
\label{ragain}
\end{equation}
Since it depends on both $H'$ and $H$, the observation of a particular value for $r$ would be helpful in constraining possible inflationary models, but by itself it would not tell us the scale for the inflationary expansion, $H$.  To do so, we would need to observe a primordial gravity wave background, since its amplitude could reveal to us the value of the Hubble scale,
\begin{equation}
\Delta^2_h = 16\pi G {H^2\over\pi^2} (-k_0\eta)^{-2\epsilon} . 
\label{gravpower}
\end{equation}
And further, if we could detect a tilt in the gravity wave spectrum, $n_T$, we would have still stronger evidence that this simple picture is correct if the tilt is found to be simply related to the scalar to tensor ratio,
\begin{equation}
n_T = - {1\over 8} r . 
\label{ntagain}
\end{equation}

The point is that even this basic picture has some fairly strong assumptions at its foundation, which we might not wish to take for granted and  which therefore must be checked.  How the universe behaves at extremely small scales can have a profound effect on each of these predictions, as a result of the same extreme expansion necessary for inflation in the first place.  What is needed to distinguish among these possibilities is a series of successively more difficult measurements.  The scalar to tensor ratio might be experimentally accessible in the relatively near future, but only by seeing the direct effects of the primordial gravity waves and measuring their spectrum's amplitude and tilt---a daunting task---can we begin to have some confidence that the minimal picture is the correct one.

In this section, we shall look at some of the different ways that the symmetry-breaking operators can alter the standard prediction for the scalar to tensor ratio.  We have chosen to use the symmetry-breaking operators since they can reproduce effects similar to those obtained by ``trans-Planckian'' modifications of the state of the inflaton, but without the need to worry about potentially uncontrolled loop corrections.  Loop corrections for states that contain nonadiabatic short-distance structures appear---from a conventional perspective \cite{emil}---to be unrenormalizable; however, this problem may be no more than an indication that, in addition to modifying the state, how we set up the quantum field theory should also be modified in a way that is consistent with the nonadiabatic state \cite{effectivestate}.  The renormalization of the symmetry-breaking operators (even with more than two powers of the field) is completely straightforward.  In the cases that we examine below, we follow each symmetry-breaking example with a similar result produced by a vacuum state with ``trans-Planckian'' structures to emphasize their similarities.

\subsection{Shifting the scalar to tensor ratio}

The simplest case is that which we encountered in de Sitter space, though we are now considering the more realistic example of a slowly rolling universe.  Consider an arbitrary set of operators with only one power of the spatial derivative, 
\begin{equation}
{\cal L}_I 
= \sum_{p=1}^\infty {c_S^{p,1}\over 3^p M^{p-1}} K^p\varphi{\cal D}\varphi
+ \sum_{p=1}^\infty {c_T^{p,1}\over 3^pM^{p-1}} K^p\tau_{ij}{\cal D}\tau^{ij} . 
\label{LIqeq1}
\end{equation}
To leading order in the initial and final times, these operators only produce ``unobservable'' shifts in the amplitudes of the power spectra, 
\begin{equation}
\Delta_\zeta{(k_0\eta)} 
= {4\pi G\over\epsilon} {H^2\over 4\pi^2} 
\biggl[ 1 + \pi \sum_{p=1}^\infty c_S^{p,1} \biggl( {H\over M} \biggr)^p 
+ \cdots \biggr]
\label{PScase1}
\end{equation}
and
\begin{equation}
\Delta_h{(k_0\eta)} 
= 16\pi G {H^2\over\pi^2} 
\biggl[ 1 + 2\pi \sum_{p=1}^\infty c_T^{p,1} \biggl( {H\over M} \biggr)^p 
+ \cdots \biggr] , 
\label{PTcase1}
\end{equation}
where we have neglected further corrections suppressed by the slow-roll parameters.  These corrections might be deemed unobservable at the level of the amplitudes of the power spectra since we could not distinguish them from an ordinary inflationary model---one without the symmetry-breaking operators---with different values for the Hubble scale and the $\epsilon$ parameter,
\begin{eqnarray}
H &\!\!\!\to\!\!\!& 
\tilde H = H \biggl[ 1 + 2\pi \sum_{p=1}^\infty c_T^{p,1} \biggl( {H\over M} \biggr)^p + \cdots \biggr]^{1/2} 
\nonumber \\
\epsilon &\!\!\!\to\!\!\!& 
\tilde\epsilon = \epsilon 
{1 + 2\pi \sum_{p=1}^\infty c_T^{p,1} \bigl( {H\over M} \bigr)^p + \cdots 
\over 
1 + \pi \sum_{p=1}^\infty c_S^{p,1} \bigl( {H\over M} \bigr)^p + \cdots } . 
\label{shifted}
\end{eqnarray}
However, their effect on the scalar to tensor ratio is not quite as invisible.  In some cases it can enhance the value of $r$, even though the slow-roll parameter $\epsilon$ remains small.

Neglecting the terms suppressed by powers of $H/M$, the scalar to tensor ratio in this model is approximately
\begin{equation}
r = 16\epsilon
{1 + 2\pi c_T^{0,1}\over 1 + \pi c_S^{0,1} }
+ {\cal O}\biggl( \epsilon {H\over M} \biggr) 
\label{epsiloncase1}
\end{equation}
In deriving this expression, we have treated the corrections perturbatively, so $c_S^{0,1}$ and $c_T^{0,1}$ are implicitly small.  Nevertheless, the case where $c_S^{0,1}$ is small and negative and $c_T^{0,1}$ is positive will modestly increase the value of $r$.

We could also leave the perturbative regime, by allowing the coefficients of these symmetry breaking terms to be large ($c_S^{0,1}, c_T^{0,1} \approx 1$) and then summing over all the graphs with an arbitrary number of insertions of the symmetry-breaking operators.  In principle, we should expect a correction of the general form, 
\begin{equation}
r = 16\epsilon f(c_S^{0,1}, c_T^{0,1}) , 
\label{epsiloncase1f}
\end{equation}
with some function $f(c_S^{0,1}, c_T^{0,1})$, though in practice it is difficult to evaluate the multiple integrals for the intermediate conformal times at which each operator is inserted.  To distinguish such very general possibilities from the minimal case requires measuring more than just the scalar to tensor ratio.  For example, the tilt of the tensor power spectrum in the general case will no longer have the simple relation of the minimal model, $r = - 8n_T$.   

What this example shows is that, depending on the details of the early universe, even for a very small value of $\epsilon$, the scalar to tensor could be observable because it is enhanced by operators or state-dependent effects.  Of course, given the difficulty of measuring the details of the primordial gravity wave spectrum, such a small value of $\epsilon$ would then mean that it would be only that much more difficult to determine whether the minimal picture does in fact hold in nature.

Before examining more interesting examples than this simple constant shift in the power spectra produced by the one-derivative operators, let us examine a similar case that occurs for the $\alpha$-states of de Sitter space \cite{alpha}.  Unlike Minkowski space, which has a unique invariant vacuum state, de Sitter space has an infinite family of invariant ``vacuum'' states.  These states are often called the $\alpha$-vacua, since they can be labeled by a single complex number $\alpha$.  The $\alpha$-states only exist in true de Sitter background so they are not strictly applicable to inflation.\footnote{In pure de Sitter space, the gravitational fluctuations decouple from the inflaton fluctuations; for example, in Eq.~(\ref{varphidef}) $\varphi = \delta\varphi + 0\times\Psi$.}  In terms of the notation introduced in Eq.~(\ref{scalarsoln}) and Eq.~(\ref{tensorsoln}), these states have $k$-independent structure functions, 
\begin{equation}
f_k=e^\alpha
\qquad 
t_k^{(r)} = e^{\tilde\alpha} . 
\label{structures}
\end{equation}
Taking the de Sitter case, $\nu=n={3\over 2}$, the power spectra are then
\begin{equation}
P_k^{\varphi,\alpha} = {H^2\over 4\pi^2} {|1-e^\alpha|^2\over 1 - e^{\alpha+\alpha^*}}
\qquad
P_k^{\tau,\tilde\alpha} = {H^2\over\pi^2} {|1-e^{\tilde\alpha}|^2\over 1 - e^{\tilde\alpha+\tilde\alpha^{*}}}
\label{alphapower}
\end{equation}
in the $-k\eta\to 0$ limit.  Here we have not assumed that the scalar and tensor modes are in the same state, $\alpha\not=\tilde\alpha$; when that is the case, their ratio can be enhanced or diminished accordingly.

The status of the $\alpha$-vacua is still not fully resolved \cite{fate}, so this last example perhaps ought to be taken more suggestively rather than too religiously.  But once again we see that ``unobservable'' shifts in the amplitudes, because it their effects on the ratio, might have observable consequences after all.

\subsection{Oscillations}

The distortions in the power spectrum produced by a general symmetry-breaking operator much more closely resemble those generated by trans-Planckian modifications of the inflaton's state, most especially those of an effective state treatment \cite{effectivestate,schalm}.  These operators introduce a $k$-dependent modulation whose amplitude also grows with the wave number, $k$.  For the scalar modes, we have in the general case,
\begin{eqnarray}
P^\zeta_k(\eta) 
&\!\!\!=\!\!\!& 
{4\pi G\over\epsilon} {H^2\over 4\pi^2} (-k\eta)^{-4\epsilon-2\delta}
\nonumber \\
&&
\biggl[ 1 + 
{c_S^{p,q}\over 2} 
\biggl( {H\over M} \biggr)^p
\biggl( {k\over k_\star} \biggr)^{q-2}
\cos\biggl( 2{M\over H}{k\over k_\star} \biggr) 
+ \cdots 
\biggr]
\nonumber \\
&&
\label{qgt2case}
\end{eqnarray}
whenever $q>2$.  For definiteness, let us choose the leading irrelevant operator ($p=0$, $q=3$); for this dimension five operator we have
\begin{eqnarray}
P^\zeta_k(\eta) 
&\!\!\!=\!\!\!& 
{4\pi G\over\epsilon} {H^2\over 4\pi^2} (-k\eta)^{-4\epsilon-2\delta}
\nonumber \\
&&
\biggl[ 1 + {1\over 2} c_S^{0,3} {k\over k_\star} 
\cos\biggl( 2{M\over H}{k\over k_\star} \biggr) 
+ \cdots 
\biggr] . 
\nonumber \\
&&
\label{p0q3case}
\end{eqnarray}

The fact that the corrections diverge as $k\gg k_\star$ does not mean that the actual power spectrum diverges, but rather that the assumption that the corrections can be described perturbatively is breaking down.  The power spectrum could remain finite beyond $k_\star$---it is simply no longer calculable in this framework.  Therefore, it is important to determine first over what range of scales this expression remains applicable.

As a minimal requirement, a mode that is just re-entering the horizon today, $k_{\rm min}$, must have been within the horizon when we start our evolution at $\eta_0$, 
\begin{equation}
{k_{\rm min}\over a(\eta_0)} \ge H(\eta_0) . 
\label{kmindef}
\end{equation}
Note again that $\eta_0$ is not necessarily the actual beginning of inflation.  In the slowly rolling limit, replacing $\eta_0$ with the $k_\star$ introduced earlier, this basic causality requirement becomes, 
\begin{equation}
{k_{\rm min}\over k_\star} \ge {H\over M} 
+ \cdots , 
\label{kmineta0}
\end{equation}
and thus the range of applicable scales is 
\begin{equation}
\biggl( {H\over M} \biggr)\, k_\star < k < k_\star . 
\label{krange}
\end{equation}
This range extends over several orders of magnitude.  The exact extent depends on the ratio between the inflationary Hubble scale $H$ and the scale of the new physics $M$, which most conservatively would be at the Planck scale, $M=M_{\rm pl}$.

Unlike other trans-Planckian corrections to the power spectrum, which are suppressed by a simple factor $H/M$, here, because the amplitude itself grows with $k$, it is not necessary for $H$ to be a significant fraction of $M$ for such effects to be observable.  Using the upper limit of the observationally allowed values of the Hubble scale, $H\sim 10^{14}$ GeV, and $M\sim M_{\rm pl}\sim 10^{18}$ GeV, the perturbative description remains applicable only over four orders of magnitude---barely enough to model corrections to the cosmic microwave background.

The description of the symmetry-breaking operators that we have been using is an effective one.  The constants $c_S^{p,q}$ and $c_T^{p,q}$ have been left general, but in any specific scenario they will assume particular values.  For example, consider a universe that contains an excited scalar field coupled to $\varphi$ which has little influence on the tensor modes $\tau_{ij}$.  In this case, we might impose the usual assumption that tensor modes are in the Bunch-Davies state, and only add the symmetry-breaking effects to the scalar field.  For the purpose of illustration, let us analyze just the leading irrelevant term $c_S^{0,3}$ which we shall set to $c_S^{0,3}=1$ to keep the expressions uncluttered.  In this case the scalar to tensor ratio will be
\begin{equation}
r = 16\epsilon\, 
{(-k_0\eta)^{2\epsilon + 2\delta} \over 
1 + {\displaystyle{1\over 2}} {\displaystyle{k_0\over k_\star}} 
\cos\biggl( {\displaystyle 2{M\over H}{k_0\over k_\star}} \biggr) } , 
\label{asyegr}
\end{equation}
evaluating $r$ at a particular value $k_0$ as usual.

In the conventional picture, the exact choice of $k_0$ is not especially important since $r$ depends only very mildly on $k_0$ in the slowly rolling limit, as $k_0^{2(\epsilon+\delta)}$.  In contrast, for the example here, where a particular symmetry-breaking operator influences the scalar modes---or where there are nonadiabatic structures in the state of the scalar field---it is possible to obtain quite different values for $r$ depending on the choice of $k_0$, especially once the amplitudes of the corrections become appreciable, $k_0\sim k_\star$.  The effect will also be enhanced if $k_0$ is near the extrema of the cosine,
\begin{equation}
k_0 \sim 2\pi n {H\over M} k_\star , 
\qquad
n\in {\mathbb Z} . 
\label{extrema}
\end{equation}
Fortunately, this sort of additional dependence, even if it exists, is not likely to have an effect on the measurements made of the microwave background.  The value chosen in the WMAP experiment, for example, is $k_0 = 2\ {\rm Gpc}^{-1}$, which corresponds to about a $2^\circ$ arc of the sky.  The prominent first acoustic peak appears only at $0.6^\circ$; so if $k_0$ were already nearing $k_\star$, then the modulation in the power spectrum would probably have significantly affected the first peak as well.  Its influence would only grow more prominent for the higher order peaks.

As mentioned, the corrections to the power spectra analyzed in this section essentially reproduce those generated in the effective state approach \cite{effectivestate,schalm} to the trans-Planckian problem of inflation.  In that picture, new short-distance structures are introduced into the state, which is also defined at an initial time $\eta_0$.  In terms of the function $f_k$ of Eq.~(\ref{scalarsoln}), these structures are 
\begin{equation}
f_k = - {1\over 4} \sum_{n=1}^\infty c_S^{(n)} {k^n\over a^n(\eta_0)M^n} . 
\label{effectivestate}
\end{equation}
The leading corrections to the power spectrum from this state are then, 
\begin{equation}
P_k^\varphi = {H^2\over 4\pi^2} \biggl[ 1 
+ {1\over 2} \sum_{n=1}^\infty c_S^{(n)} \biggl( {k\over k_\star} \biggr)^n 
\cos\biggl( 2 {M\over H} {k\over k_\star} \biggr) \cdots \biggr] , 
\label{powereff}
\end{equation}
where we have again neglected the dependence on the slow-roll parameters.  This expression matches exactly with the leading result produced by the operator with the maximum number of spatial derivatives at a given dimension; that is, $p=0$ and $q=d-2$.

We conclude this subsection with a few more general remarks.  As we have seen, the effect of the generic symmetry-breaking operators, those with more than two spatial derivatives, is a $k$-dependent modulation whose amplitude grows with large values of $k$, or equivalently, at large multipole moments.  Depending on the number of time derivatives in the operator, the effect may or may not be suppressed by extra powers of $H/M$; but because of the growth of the amplitude with $k$, the ratio between the dynamical scale of inflation, $H$, and the scale associated with the new phenomena, $M$, does not need to be comparatively ``large''---meaning in this instance something of the order of $10^{-2}$ to $10^{-3}$---to have an observable effect.  

The role of $H/M$ for such effects is more subtle than if the correction were simply suppressed by this ratio.  Here it sets the range of wave numbers over which the effective theory treatment of the modes remains valid---the lower end is set by the requirement that inflation should provide a causal explanation of the origin of the inhomogeneities of the universe ($k>aH$) while the upper end is essentially the trans-Planckian limit, the fact that we do not want to extend our theory into the regime where quantum fluctuations are shorter than the Planck length, ($k<aM_{\rm pl}$), or some other scale where new phenomena might occur ($M\le M_{\rm pl}$).  Of course, since inflation exists in a time-dependent background, both of these statements are themselves time-dependent.  Therefore we have defined our states at an initial time $\eta_0$.  

The physical meaning of $\eta_0$ depends to an extent on our intention for the effective theory.  If we are modeling a specific scenario, such as the example above of an excited field coupled to the scalar field, then $\eta_0$ might be associated with a physical event, for example the last time that the excited field influences the scalar field.  It could correspond to the actual beginning of inflation as well, although in this case $\eta_0$ cannot be much beyond the minimal amount needed to address the horizon problem since the trans-Planckian bound will similarly move earlier too.  Finally, $\eta_0$ might not correspond to any real physical scale at all, but only to a cut-off as in an ordinary regularized field theory in flat space.  Physical results would therefore not depend on $\eta_0$, so there would need to be some form of dimensional transmutation which replaces $\eta_0$ with a parameter labeling some physical class of states.

Although the restriction to the scales $(H/M)k_\star < k < k_\star$ might appear to be a limitation only of our effective approach, it is not.  {\it Any\/} treatment of inflation that defines its modes above $k_\star$ is---often without explicitly stating so---making some assumption about the behavior of {\it quantum gravitational\/} fluctuations at lengths smaller than a Planck length.  Conversely, even if the standard Bunch-Davies picture is finally confirmed, we would have learned something profound about nature at small scales and its essential flatness.  This property would then need to be incorporated into any theory seeking a quantum description of gravity.

\subsection{Oscillations at large scales}

As a final example, we shall look at a case where a power spectrum is modified at {\it long\/} distances.  Effective approaches are usually applied to short distances, so this section lies a little outside the main approach of this article.  However, since the observed scalar power spectrum does show some evidence \cite{silk} of being suppressed at long wavelengths, it would be quite useful to see how a general, model-independent framework might be applied to these scales.

The WMAP constraints on the primordial power spectra are almost always described in the context of the standard inflationary model.  In this picture, the spectra are largely featureless, with only an amplitude and a small tilt.\footnote{In principle, there is a parameter too for the running of the spectral index, $dn_S/d\ln k$, but it has not been measured yet.}  If instead we make no initial assumptions about what produced the primordial fluctuations, what we can infer from the actual data looks rather different.  For example, the primordial spectrum extracted from the cosmic microwave background radiation in \cite{silk} has noticeable oscillations, together with a large, gradual suppression of power at longer wavelengths.  Admittedly, some of these features appear where the cosmic variance is large, so it may never be possible to say whether there is some underlying mechanism causing this suppression or it is merely a random artifact.  Still, the value of $k_0$ used to define $r$ is usually chosen in this same region, so if this suppression is a real feature of the primordial perturbations and if the mechanism causing it affects the scalar and tensor fluctuations differently, their ratio might again differ from the usual slow-roll inflationary prediction.

In the effective descriptions of the initial state introduced in \cite{effectivestate}, it is possible to include modifications of the state at large wavelengths as well as at small.  In an effective theory, these structures require {\it relevant\/} boundary operators.  For example, if we introduce a new scale $\mu$ associated with these long-distance structures in the state, then for wavenumbers $k>a\mu$, the following structure would produce the leading effect
\begin{equation}
f_k = ic {a(\eta_0)\mu\over k} + \cdots . 
\label{largefk}
\end{equation}
As in the case of the trans-Planckian corrections, where we introduced a wavenumber $k_\star$ whose wavelength is equal to $1/M$ at the initial time $\eta_0$, we can similarly define a wavenumber associated with this scale $\mu$, 
\begin{equation}
k_L \equiv a(\eta_0)\mu = {\mu\over M} k_\star , 
\label{kLdef}
\end{equation}
neglecting slow-roll corrections.  Clearly, to avoid trans-Planckian modes appearing simultaneously with these effects, we should have distinct scales, with $\mu\ll M$.  The relative size of $\mu$ to $H$, however, depends on our purpose.  If we would like the physics associated with these new structures to lie within the same causal horizon, then we should choose $H<\mu<M$, which for a relatively low scale of inflation would still easily provide a sufficient range to cover the wavenumbers responsible for the fluctuations observed in the microwave background while avoiding the trans-Planckian regime.  Alternatively, we might wish to model more general structures, even beyond the horizon $\mu\le H$, in much the same way as we allow for general trans-Planckian structures.  The goal in this instance is not to explain how they arose so much as it is to place bounds on their possible observational effects.

\begin{figure}[!tbp]
\includegraphics{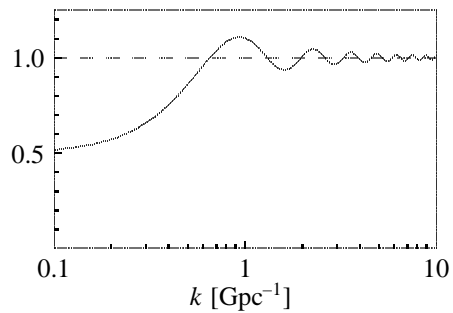}
\caption{The suppression of the power spectrum of $\zeta$ produced by term $f_k=ick_L/k$ in its state, relative to the standard result and ignoring small corrections in the slowly rolling limit.  The values of $c$ and $k_L$ chosen for this graph are those given in Eq~(\ref{WMAPvals}).\label{suppress}}
\end{figure}

Treating these effects perturbatively, the leading influence---for example---on the scalar power spectrum would be 
\begin{equation}
P_k^\varphi(\eta) = {H^2\over 4\pi} 
\biggl[ 1 - 2c {k_L\over k} \sin \biggl( 2 {\mu\over H} {k\over k_L} \biggr) + \cdots \biggr] . 
\label{largescale}
\end{equation}
To be applicable to the suppression at long wavelengths, we would very roughly need
\begin{eqnarray}
{1\over k_L} &\!\!\!\sim\!\!\!& 2.4\ {\rm Gpc}\, \cdot {H\over\mu}
\nonumber \\
c &\!\!\!\sim\!\!\!& {1\over 8} {H\over\mu} , 
\label{WMAPvals}
\end{eqnarray}
which we have expressed in terms of the dimensionless ratio $H/\mu$.  The effect of this correction for these values, as a fraction of the standard result, is thus
\begin{equation}
P_k^\zeta(\eta) \bigg/ \biggl( {4\pi G\over\epsilon} {H^2\over 4\pi^2} \biggr) 
= 1 - {\sin \bigl( 4.8k\ {\rm Gpc} \bigr)\over 9.6k\ {\rm Gpc}} + \cdots
\label{diminish}
\end{equation}
and is shown in the Fig~\ref{suppress}.

Since this effect is a relevant one with respect to the effective theory, its importance diminishes at short distances and grows at long ones.  As before, the fact that the corrections are growing large at scales $k<k_L$ signals the breaking down of the perturbative treatment and that at such scales potentially any effect scaling as a positive power of $k_L/k$ will also need to be included.  Of course, it never makes sense to go to the absolute limit $k\to 0$, since whatever lies beyond the horizon today---roughly of the order of $0.1$ ${\rm Gpc}^{-1}$---is completely inaccessible to us.

The physical meaning of the new scale $k_L$ is not explained by an effective theory treatment and requires some new principle to account for it.  While it could happen that $k_L$ is related to something happening during inflation, that this scale would be precisely the right size to be important now, when the universe is so many orders of magnitude older, seems highly unlikely.  A more natural interpretation is that this scale is due to some modification in how the universe evolves, which only becomes apparent on scales of the order of the horizon size.  The reason, in this case, that it appears unnatural is simply that we have mapped an effect whose origin lies in a modification of the dynamics today back to a time deep within the inflationary era, which was done so that the origin and the suppression of the primordial fluctuations are treated on a common footing.

It would be good to develop this large-scale effective approach further \cite{progress}.  In the absence of any compelling models to describe how the universe evolves at very large scales, it is helpful to formulate a framework for classifying the general possible modifications that could occur.  So far, the experimental constraints on the primordial power spectrum, inferred from measurements of the power spectrum of the temperature fluctuations in the microwave background, are still very weak.  However, these constraints will continually be improving,  beginning with the measurements by the Planck satellite and then later with a variety of experiments designed to study the large scale structure through the 21 cm line for the hyperfine transition in hydrogen.  Usually the form for primordial perturbations used in fits of such data is that arising from inflation.  Yet, this spectrum should also emerge naturally, without imposing its weak power-law behavior from the beginning, if we are to have some confidence that the inflationary picture could be the correct one for our universe.

\section{Conclusions}

In this article we have examined some of the ways that a breakdown of local Lorentz-invariance at very short distances would affect the scalar to tensor ratio predicted by inflation.  We chose this model since it reproduces many of the features---and some new ones as well---seen in effective state treatments of the trans-Planckian problem, but in a slightly tamer setting.  One of our reasons for looking specifically at $r$ is that it subtly influences the inferences we might make about inflation from observations.  To take one example, the amplitude of the scalar fluctuations is sometimes used indirectly to constrain the inflationary scale $H$.  Unlike the tensor fluctuations, its amplitude depends on two unknown quantities:  both $H$ and the slow-roll parameter $\epsilon$,
\begin{equation}
\Delta_\zeta^2 
\approx {1\over 8\pi^2} {1\over\epsilon} {H^2\over M_{\rm pl}^2} .  
\label{zetaHMpl}
\end{equation}
If we use the standard relation, $r = 16\epsilon$, derived earlier, then the empirical bound on $r$ together with the measured value for the amplitude $\Delta_\zeta^2$ \cite{wmap}, 
\begin{equation}
\Delta_\zeta^2({\rm exp}) = (2.457^{+0.092}_{-0.093}) \times 10^{-9} , 
\label{zetaEXP}
\end{equation}
places an upper limit on the scale $H$, 
\begin{equation}
H \sim \pi M_{\rm pl} \sqrt{\Delta_\zeta^2} \sqrt{ {r\over 2} }
\sim 2.6\sqrt{r} \times 10^{14}\ {\rm GeV} . 
\label{zetaasr}
\end{equation}
From what we have seen, trans-Planckian corrections might change this relation---for example, by reducing the value of $r$ as a factor of $\epsilon$, which would allow for a higher value of $H$.

The simple symmetry-breaking operators that we have studied can affect $r$ in a variety of ways, though the most characteristic ones are those that oscillate about the standard prediction for $P_k^\zeta$ (and $P_k^h$) with an amplitude that grows as a power of scale $k$.  It is important again to emphasize that we have been implicitly using an effective treatment, so that when $k$ becomes comparable or even much larger than the trans-Planckian threshold $k_\star$, the new effects are not necessarily diverging.  Instead, the perturbative treatment is breaking down, and all powers of $k/k_\star$ are becoming equally important.  Note too that it is often thought that trans-Planckian effects are suppressed by $H/M$.  Certainly this happens in some classes of theories, and in some of the effects studied here, but it is not invariably so.  The generic corrections that we have examined scale instead as a power of $k/k_\star$.  Thus, even if the inflationary scale is tiny compared to the scale $M$, trans-Planckian effects are not necessarily negligible.  What is more important is how close the physically relevant $k$'s are to the trans-Planckian threshold $k_\star$.

The main lesson from this and other works on the trans-Planckian problem is that without some understanding of the quantum behavior of gravity, the only safe scales $k$ that are relevant to our observable universe are those that lie within the range
\begin{equation}
{H\over M} k_\star < k < k_\star . 
\label{rangeagain}
\end{equation}
If the lower end corresponds to modes that still have not yet reentered the horizon, then the range available to the scales relevant to what we observe in the cosmic microwave background will be still narrower, or it may not even exist at all.  These constraints do not apply just to the examples described here, where the inapplicability of the perturbative description as we approach $k\sim k_\star$ is merely clearer, but to any theory of inflation.  Inflationary models invariably must make some assumption about how nature behaves over scales shorter than a Planck length.  Sometimes these assumptions are guided by what is reasonable, or renormalizable, for a quantum field theory in a curved background.  However, in inflation, we are speaking of the fluctuations of the background itself---and what might be reasonable there is an altogether different matter.

\begin{acknowledgments}

\noindent
This work was supported in part by the EU FP6 Marie Curie Research and Training Network ``UniverseNet'' (MRTN-CT-2006-035863) and by the Niels Bohr International Academy.  I would like to thank Rich Holman for his comments on this article, as well as Anupam Mazumdar for initially suggesting this problem.

\end{acknowledgments}

\end{document}